# Magneto-photoluminescence of exciton Rydberg states in monolayer $WSe_2$

Erfu Liu[1], Jeremiah van Baren[1], Takashi Taniguchi[2], Kenji Watanabe[2],
Yia-Chung Chang[3], Chun Hung Lui[1*]

[1] Department of Physics and Astronomy, University of California, Riverside, CA 92521, USA.

[2] National Institute for Materials Science, Tsukuba, Ibaraki 305-004, Japan

[3] Research Center for Applied Sciences, Academia Sinica, Taipei 11529, Taiwan

[*] Corresponding author. Email: joshua.lui@ucr.edu

Abstract: Monolayer $WSe_2$ hosts a series of exciton Rydberg states denoted by the principal quantum number $n$ = 1, 2, 3, etc. While most research focuses on their absorption properties, their optical emission is also important but much less studied. Here we measure the photoluminescence from the 1s – 5s exciton Rydberg states in ultraclean monolayer $WSe_2$ encapsulated by boron nitride under magnetic fields from –31 T to 31 T. The exciton Rydberg states exhibit similar Zeeman shifts but distinct diamagnetic shifts from each other. From their luminescence spectra, Zeeman and diamagnetic shifts, we deduce the binding energies, g-factors and radii of the 1s – 4s exciton states. Our results are consistent with theoretical predictions and results from prior magneto-reflection experiments.

Key words: Magneto-photoluminescence, exciton Rydberg states, monolayer $WSe_2$, Zeeman shift, diamagnetic shift.

Monolayer transition metal dichalcogenides (TMDs), such as $MoS_2$ and $WSe_2$, possess strong Coulomb interactions due to the reduced dielectric screening and strong spatial confinement in strictly two-dimensional (2D) systems[1-3]. Their electrons and holes can form excitons with binding energies 300 – 600 meV, an order of magnitude larger than those in traditional quasi-2D quantum well systems[4-16]. The strong interaction produces a series of internal exciton states. They mimic the Rydberg states in the hydrogen atom and can likewise be characterized by the principal quantum number ($n$ = 1, 2, 3 …) and the *s*, *p*, *d* orbitals with quantized angular momentum (Figure 1a)[6-9, 17-20]. The exciton Rydberg states in monolayer TMDs possess many remarkable properties, such as nonhydrogenic energy spacing[6], nonlocal screening[17], superior valley polarization and coherence[21], strong two-photon coupling[8, 9, 22], and ultrafast photo-response[23]. These distinctive excitonic properties are intriguing topics for both fundamental research and novel optical and electronic applications.

The exciton Rydberg states in monolayer TMDs have been much studied with absorption-related spectroscopy[6, 7, 9, 18, 24]. In particular, by combining high magnetic field, reflection spectroscopy can reveal the oscillator strength, energy levels, orbital size, and magnetic energy shifts of TMD excitons[24-30]. Nonetheless, prior research mainly focuses on the absorption properties of the exciton Rydberg states. The other aspect of the Rydberg states – the optical emission – is also important, but much less studied experimentally. This is because high-lying exciton levels have short lifetime and weak oscillator strength, leading to tiny optical emission. Fabrication of ultraclean samples and sensitive measurement are necessary to reveal the fine optical features from the high-lying exciton levels[21]. As magnetic field can strongly modify the exciton properties, it is particularly interesting to study the optical emission of exciton Rydberg states under strong magnetic field. Such experimental studies, however, have been lacking thus far.

In this Letter, we measure the photoluminescence (PL) of exciton Rydberg states in ultraclean monolayer $WSe_2$ encapsulated by boron nitride (BN) under magnetic fields from B = –31 T to 31 T. We observe light emission from five (1s – 5s) exciton Rydberg states. From the PL spectra, we can accurately extract the Zeeman shifts and diamagnetic shifts of the first four states (1s – 4s) and further deduce their g-factors, exciton radii and binding energies. The results can be quantitatively simulated by a model calculation based on non-hydrogenic excitonic interaction and non-local screening in an embedded 2D system.

Our magneto-PL experiment complements and extends prior magneto-reflection experiments, which focus on the absorption properties of the TMD excitons. Our PL results also provide important information to elucidate the diverse properties of exciton Rydberg states in 2D materials.

We conducted the experiment in a 31-Tesla cryogenic magneto-optical system in the National High Magnetic Field Laboratory in Tallahassee, Florida. We measured the PL from BN-encapsulated monolayer $WSe_2$ at temperature near T = 4 K (Figure 1b). The samples were excited by a 532-nm continuous laser through an optical fiber. Relatively high incident laser power (~1 mW) was applied to reveal the weak PL signals of high-lying exciton states. The PL was collected in a back-scattering geometry. We only detected the PL with left-handed circular polarization, which corresponds to the K-valley emission in monolayer $WSe_2$ (see Supporting Information for details).

Figure 1c displays a representative PL spectrum of monolayer $WSe_2$ with no magnetic field. Our ultraclean samples exhibit remarkably sharp PL lines. We can identify the 1s, 2s, 3s and 4s states of A exciton, which have PL energies ($E_{PL}$) of 1.7119, 1.8425, 1.8638, and 1.8727 eV, respectively (Table 1). From these data we extrapolate a free-particle band gap of $E_g$ = 1.884 eV by using a quantitative model, which will be described later. The thus obtained binding energies ($E_b = E_g - E_{PL}$) of 1s – 4s excitons are 172.1, 41.5, 20.2 and 11.3 eV, respectively (Table 1). These binding energies are comparable to those of prior reflection experiments on similar BN-encapsulated monolayer $WSe_2$ samples (see comparison in Figure 4)[24]. We only consider the A exciton in this paper. The B exciton of $WSe_2$, being ~400 meV above, is well separated from the Rydberg series of A exciton.

To further explore the A-exciton Rydberg states, we apply strong out-of-plane magnetic field on our samples. Figure 2a, c displays the PL maps and cross-cut spectra of monolayer $WSe_2$ at B = –31 to 31 T. The application of magnetic field significantly shifts the energies of all exciton peaks. The high-lying states shift much more than the low-lying states. The resultant enlarged separation between adjacent peaks helps us distinguish different exciton states. In particular, the 4s state, which appears as a shoulder of the 3s peak at zero field, becomes isolated and well recognized in high magnetic field. To further enhance the weak features, we take the second energy derivative of the PL intensity ($d^2I/dE^2$), in which the exciton states appear as dips (Figure 2b, d). In the second-order differential PL maps, all exciton features are sharpened and their energy shifts become more traceable. Remarkably, a new PL feature appears above the 4s state at B > 5 T. This feature is weak but discernable; it corresponds to the 5s state (Figure 2b). Besides, we observe a small PL feature at 1.85 – 1.86 eV above the 2s state (marked by the open dots in Figure 2c-d). It shifts parallelly with the 2s state with the magnetic field. The origin of this new PL feature is unknown and requires further investigation.

Figure 3a displays the energy of all exciton PL features as a function of magnetic field. According to prior research, the magnetic-field-dependent energy shift consists of two components – the valley Zeeman shift[31-36] and the diamagnetic shift[24, 25]. The Zeeman shift is an odd function of B field. By taking the difference between the energies at opposite B fields, we obtain the Zeeman splitting energy (Figure 3b). By contrast, the diamagnetic shift is an even function of B field. By taking the average of the energies at opposite B fields, we obtain the diamagnetic shift (Figure 3c). We have extracted the Zeeman splitting energies and diamagnetic shifts of the 1s – 4s states from our PL data (Figure 3b, c). The corresponding shifts of the 5s peak cannot be extracted accurately due to its weak PL signal.

We first consider the Zeeman effect. Prior research has shown that out-of-plane magnetic field can lift the valley degeneracy in monolayer TMDs[31-33, 36]. Due to the opposite spin and orbital configurations of the two time-reversal valleys, the magnetic field can enlarge the band gap at one valley but diminish the band gap at the other valley. The difference between the energy gaps of the two valleys is the valley Zeeman (ZM) splitting energy, which can be expressed as:

$$\Delta E_{ZM} = g\mu_B B \qquad (1)$$

Here g is the effective g-factor and $\mu_B = 5.788 \times 10^{-5}$ eV/T is the Bohr magneton. As the Zeeman shift is an odd function of B field, the Zeeman splitting energy can be obtained as the difference between the valley PL energies at opposite B fields. For the bright A exciton with spin singlet, the spins contribute little to the exciton Zeeman shift. The major contribution comes from the atomic orbitals in the valence band, which has opposite azimuthal quantum numbers $m = +2$ and $–2$ and opposite magnetic moment at the two valleys[31-33, 36]. They contribute an effective g-factor of –4 to the valley Zeeman splitting. In our experiment, the Zeeman splitting energies of 1s – 4s states all exhibit linear dependence with g-factors between – 4.0 and – 4.3 (Figure 3b, Table 1). The results are consistent with the above prediction as well as results from prior reflection experiments[31-33, 36].

We next consider the diamagnetic shift. In the weak-field limit, where the characteristic magnetic length $l_B = \sqrt{\hbar/eB}$ exceeds the exciton radius and the Landau level spacing is much smaller than the exciton binding energy, the diamagnetic (DM) shift of an exciton can be expressed as[24-26]:

$$\Delta E_{DM} = \frac{e^2}{8\mu}\langle r^2 \rangle B^2 = \sigma B^2 \tag{2}$$

Here $\mu = (m_e^{-1} + m_h^{-1})^{-1}$ is the exciton's reduced mass; $\sigma$ is the diamagnetic coefficient; $r$ is the radial coordinate of the exciton; $\langle r^2 \rangle = \langle \psi | r^2 | \psi \rangle$ is the expectation value of $r^2$ over the exciton envelope wave function $\psi$. The exciton's root-mean-square radius is $r = \sqrt{\langle r^2 \rangle} = \sqrt{8\mu\sigma}/e$. A larger diamagnetic shift indicates a larger exciton size. The larger high-lying excitons have much larger diamagnetic shifts than the smaller low-lying excitons.

In the strong-field limit, where $l_B$ is much smaller than the exciton radius and the Landau level spacing exceeds the exciton binding energy, the optical transition mainly occurs between the Landau levels in the valence and conduction bands. In this regime, the transition energy increases approximately linearly with B as $\left(N + \frac{1}{2}\right)\hbar\omega_c$ for all exciton states ($\omega_c = \hbar eB/\mu$ is the exciton cyclotron energy). In the regime of mediate B field, the exciton energy will gradually transit from $B^2$ to B dependence[24, 25, 28-30].

Figure 3c displays the diamagnetic shifts of 1s – 4s exciton states. The 1s and 2s energies both exhibit $B^2$ dependence in the whole measured range (B = 0 – 31 T). B = 31 T is still a weak field for the 1s and 2s states because of their large binding energies. By using quadratic fits, we extract their diamagnetic coefficients to be $\sigma_{1s}$ = 0.24 μeV/T$^2$ and $\sigma_{2s}$ = 6.17 μeV/T$^2$. Previous research has determined that the reduced mass of A exciton in monolayer WSe$_2$ is $\mu$ = 0.20 $m_e$[24], where $m_e$ is the free electron mass. From the relationship $r = \sqrt{8\mu\sigma}/e$, we calculate the exciton root-mean-square radii to be $r_{1s}$ = 1.58 nm and $r_{2s}$ = 8.09 nm for the 1s and 2s states (Table 1).

The 3s and 4s states exhibit $B^2$ dependence only at B < 10 T and gradually transit to linear B-field dependence at B > 10 T. B = 31 T is an intermediate field for the 3s and 4s states because of their relatively small binding energies. By quadratic fits on the low-field data (lines in Figure 3c), we extract their diamagnetic coefficients to be $\sigma_{3s}$ = 28.4 µeV/T$^2$ and $\sigma_{4s}$ = 57.8 µeV/T$^2$. The corresponding exciton radii are $r_{3s}$ = 17.35 nm and $r_{4s}$ = 24.75 nm. Table 1 and Figure 4 summarize the extracted exciton radii of 1s – 4s states. Our results agree with prior magneto-reflection experiments, which measured the exciton radii of 1s – 3s states (no 4s state) in monolayer WSe$_2$[24].

To quantitatively explain our data, we have carried out comprehensive model calculations with effective masses extracted from the density functional theory (DFT). In the calculation, we use the non-hydrogenic Keldysh potential for excitonic interactions in 2D materials[6, 16, 37-43]:

$$V(r) = -\frac{e^2}{8\varepsilon_0 r_0}\left[H_0\left(\frac{\kappa r}{r_0}\right) - N_0\left(\frac{\kappa r}{r_0}\right)\right] \quad (3)$$

Here $H_0$ and $N_0$ are the Struve and Neuman functions of zeroth order, respectively; $r_0$ is an effective screening length of monolayer WSe$_2$; $\kappa$ is the dielectric constant of the encapsulating BN[44]. $V(r)$ scales as $1/\kappa r$ when $r \gg r_0$ but diverges only weakly as $\log(r)$ when $r \ll r_0$ due to the increased screening from the WSe$_2$ layer. We obtain the best-fit parameters $r_0$ = 5 nm and $\kappa$ = 3.97 by adjusting their values within the physical range to fit the energies of the observed exciton states. Our calculation predicts a free-particle band gap of 1.884 eV for monolayer WSe$_2$. Figure 4a-b display our calculated recombination energies and radii of the 1s – 4s exciton states. They agree excellently with the experimental results. We have also calculated the diamagnetic shifts of the 1s – 5s states (Figure 3d). The calculation quantitatively reproduces the observed diamagnetic shifts in experiment (see Supporting Information for details).

In summary, we have investigated the exciton Rydberg series of ultraclean monolayer WSe$_2$ by magnetic-field-dependent photoluminescence (PL). We observe the PL from the 1s – 5s exciton states. We extract the binding energies, g-factor and radii of the 1s – 4s excitons from their PL spectra, Zeeman and diamagnetic shifts. Compared to prior reflection experiments that involve optical interference in stacked materials on the Si/SiO$_2$ substrate, our magneto-PL experiment and analysis are much more straightforward to implement. Moreover, our PL measurement appears to be more sensitive than previous experiments. We can reveal the 5s state and the full range of diamagnetic shifts of the 4s state, which were not observed in prior magneto-reflection experiments. Besides, the PL experiment can reveal the valley polarization of the exciton Rydberg states (see Supporting Information). Overall, our research demonstrates magneto-PL to be an efficient and powerful method to investigate the exciton Rydberg states in 2D semiconductors. Our PL results will also provide key information for novel optical and opto-electronic applications of TMD materials.

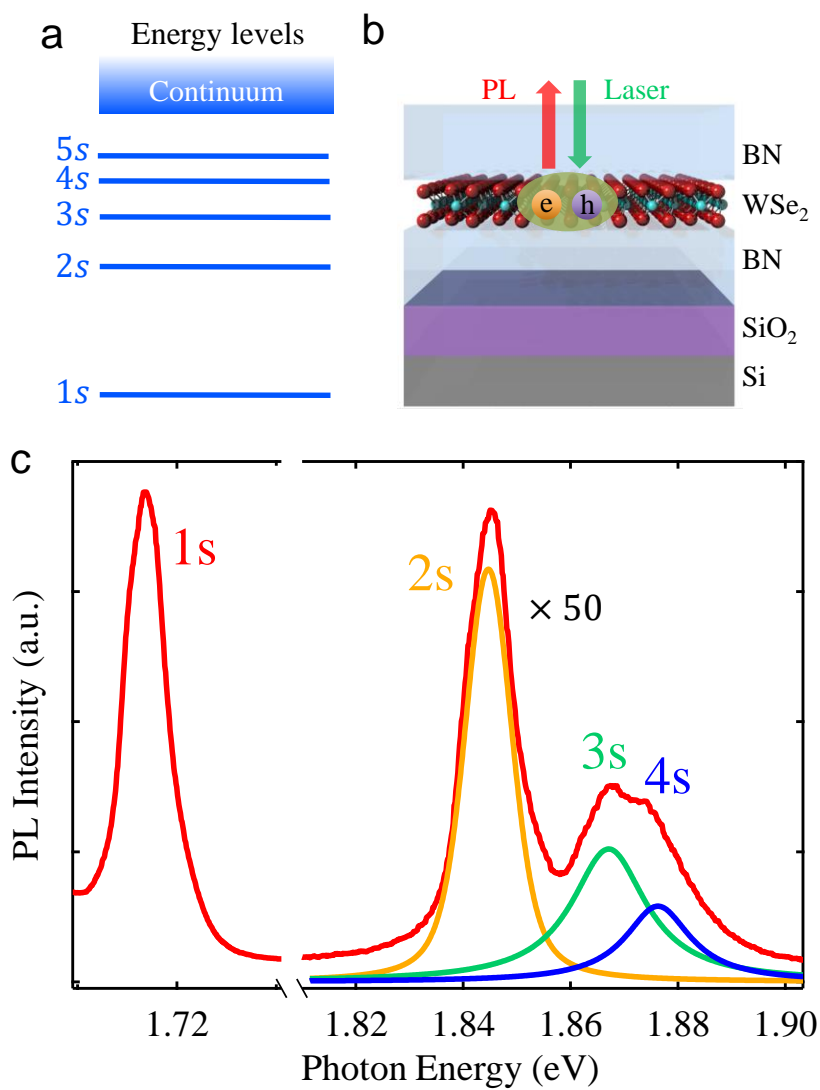

**Figure 1.** Photoluminescence (PL) of the A-exciton Rydberg states in monolayer $WSe_2$. (a) Schematic internal energy levels of an exciton in 2D semiconductors. (b) Schematic of our PL experiment on monolayer $WSe_2$ encapsulated by boron nitride flakes. (c) The PL spectrum of monolayer $WSe_2$ under 532-nm continuous laser excitation at temperature T = 4 K with no magnetic field. The spectrum at 1.81-1.90 eV is multiplied for 50 times for clarity. The red line is the experimental spectrum; the orange, green and blue lines are Lorentzian fits to reveal the 2s, 3s and 4s exciton states.

Fig. 2

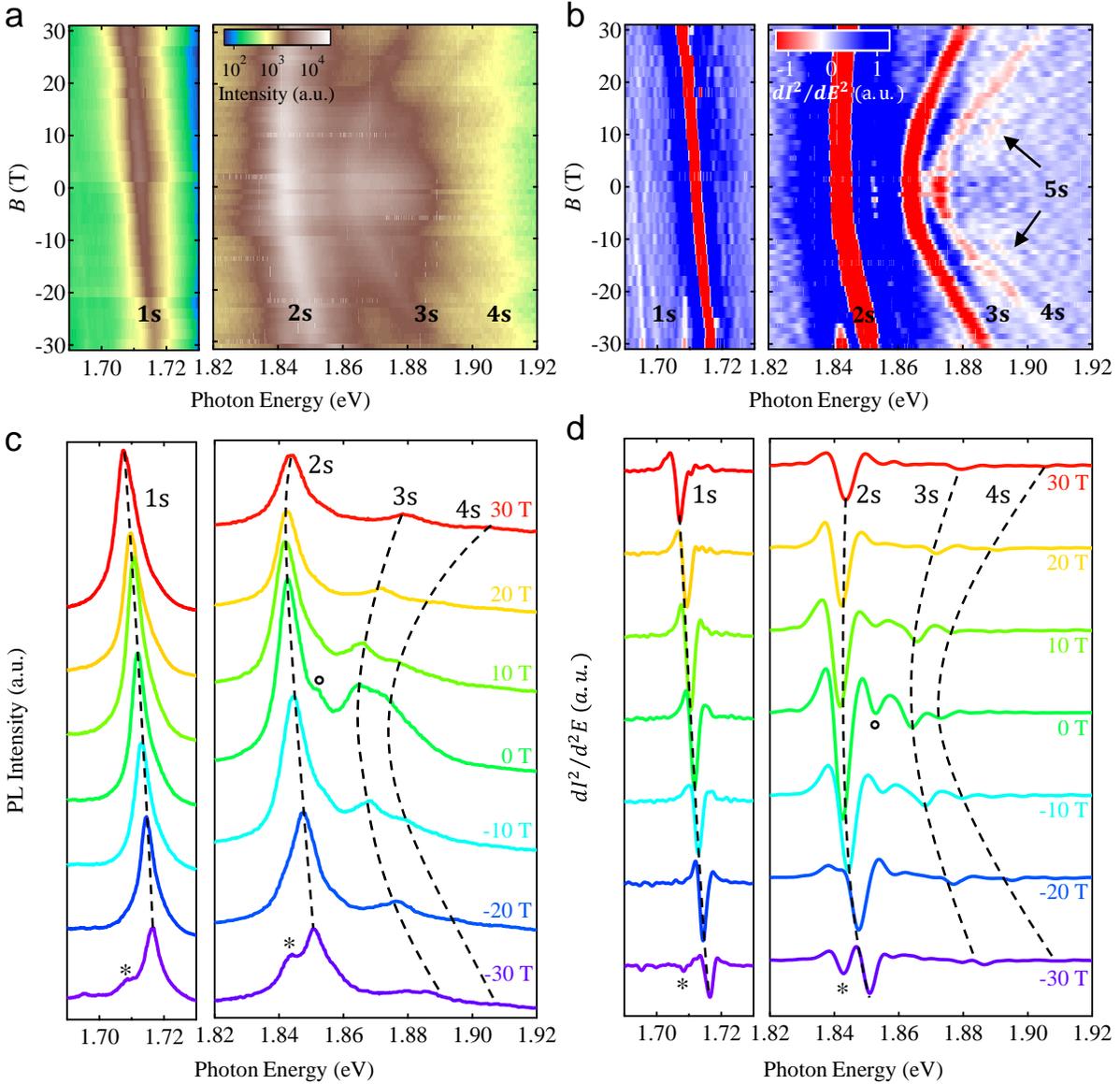

**Figure 2.** Photoluminescence (PL) of A-exciton Rydberg states in monolayer WSe$_2$ under magnetic field. (a) The logarithmic PL maps at magnetic fields B = -30 to 30 T. (b) The color map of the second energy derivative of PL intensity ($d^2I/dE^2$) in panel a. (c) The cross-cut PL spectra from panel a at selective magnetic fields. (d) The cross-cut $d^2I/dE^2$ spectra from panel b at selective magnetic fields. The dashed lines highlight the shift of Rydberg states. The PL features marked by the * symbol at B = –30 T come from the other valley due to imperfect helicity selection in the measurement.

Fig. 3

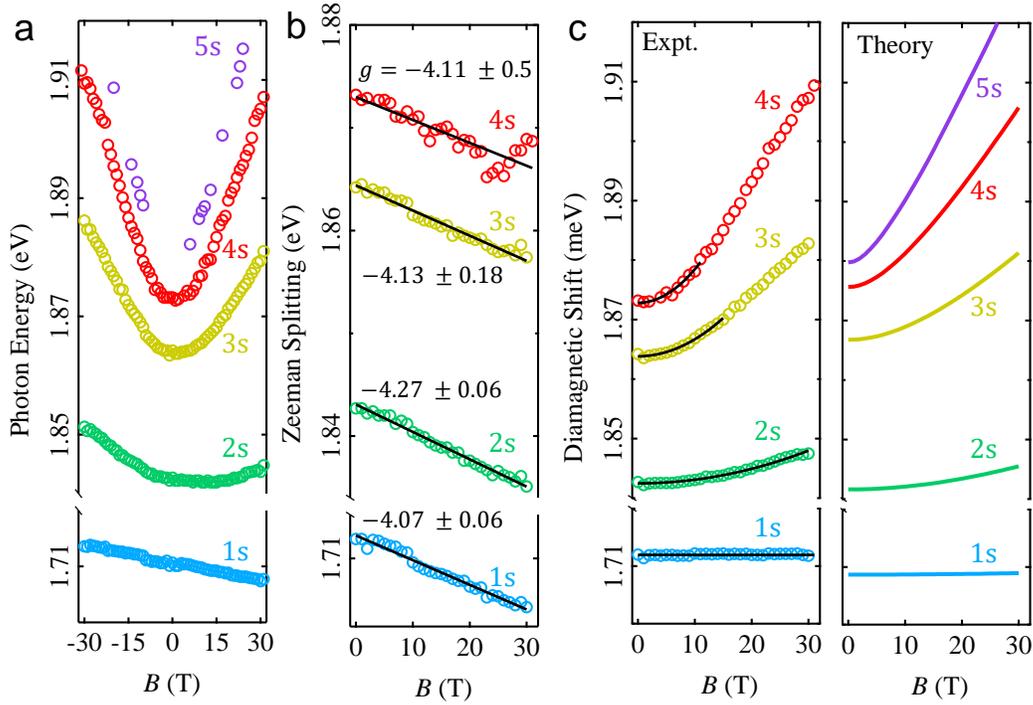

**Figure 3.** The energy shifts of A-exciton Rydberg states in monolayer WSe$_2$ under magnetic field. (a) The exciton PL energy as a function of magnetic field, as extracted from the data in Figure 2. (b) Zeeman shifts extracted from panel a. The g-factors are obtained from linear fits of the data. (c) Diamagnetic shifts extracted from panel a. The lines are quadratic fits. (d) Predicted exciton diamagnetic shifts as described in the text.

Fig. 4

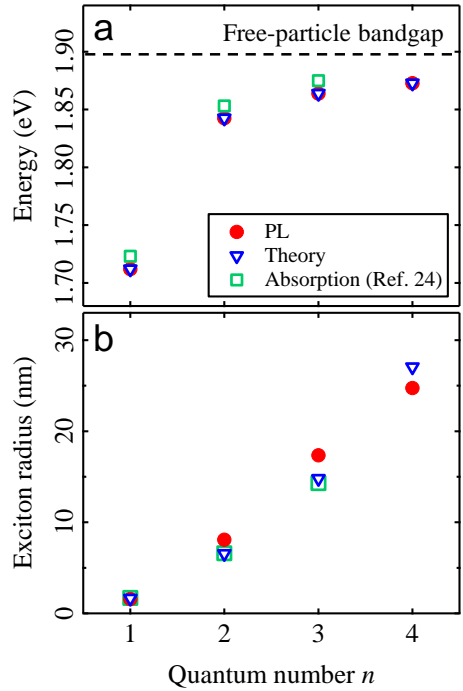

**Figure 4.** The energy and radii of Rydberg excitons in monolayer $WSe_2$. (a) The recombination energy and (b) radii of A-exciton states (red dots) extracted from the PL data in Figure 3, in comparison with the theoretical predictions (blue triangles) and results from a prior reflection experiment (green squares; Ref. 24). The dashed line denotes the predicted free-particle band gap by our model.

**Table 1.** The PL energy ($E_{PL}$), binding energy ($E_b$), g-factor, diamagnetic coefficient ($\sigma$) and root-mean-square radius ($r$) of the exciton Rydberg states in BN-encapsulated monolayer WSe$_2$. Our theoretical values of $E_b$ are shown in parentheses.

|    | $E_{PL}$ (eV) | $E_b$ (meV)  | g     | $\sigma$ (μeV/T²) | $r$ (nm) |
|----|---------------|--------------|-------|-------------------|----------|
| 1s | 1.7119        | 172.1 (172.1)| − 4.07| 0.24              | 1.58     |
| 2s | 1.8425        | 41.5 (43.8)  | − 4.27| 6.17              | 8.09     |
| 3s | 1.8638        | 20.2 (19.5)  | − 4.13| 28.4              | 17.35    |
| 4s | 1.8727        | 11.3 (11.0)  | − 4.11| 57.8              | 24.75    |

**Corresponding Author**

*Email: joshua.lui@ucr.edu

**Author Contributions**

E.L and J.v.B fabricated the device and carried out the experiment. T.T. and K.W. provided the BN crystals. Y.C.C. did the theoretical calculations. C.H.L. supervised the project. E.L. and C.H.L. wrote the manuscript with input from other authors.

**Funding**

Y.C.C. is supported by Ministry of Science and Technology, R.O.C. under grant no. MOST 107-2112-M-001-032. K.W. and T.T. acknowledge support from the Elemental Strategy Initiative conducted by the MEXT, Japan and the CREST (JPMJCR15F3), JST.

**Notes**

The authors declare no competing financial interest.

**ACKNOWLEDGMENTS**

We thank Dmitry Smirnov and Zhengguang Lu for assistance in the magneto-optical experiment.

*Supporting Information for*

**Magneto-photoluminescence of exciton Rydberg states in monolayer WSe$_2$**

Erfu Liu[1], Jeremiah van Baren[1], Takashi Taniguchi[2], Kenji Watanabe[2],

Yia-Chung Chang[3], Chun Hung Lui[1*]

[1] Department of Physics and Astronomy, University of California, Riverside, CA 92521, USA

[2] National Institute for Materials Science, Tsukuba, Ibaraki 305-004, Japan

[3] Research Center for Applied Sciences, Academia Sinica, Taipei 11529, Taiwan

[*] Corresponding author. Email: joshua.lui@ucr.edu

## 1. Experimental setup for magneto-optical spectroscopy

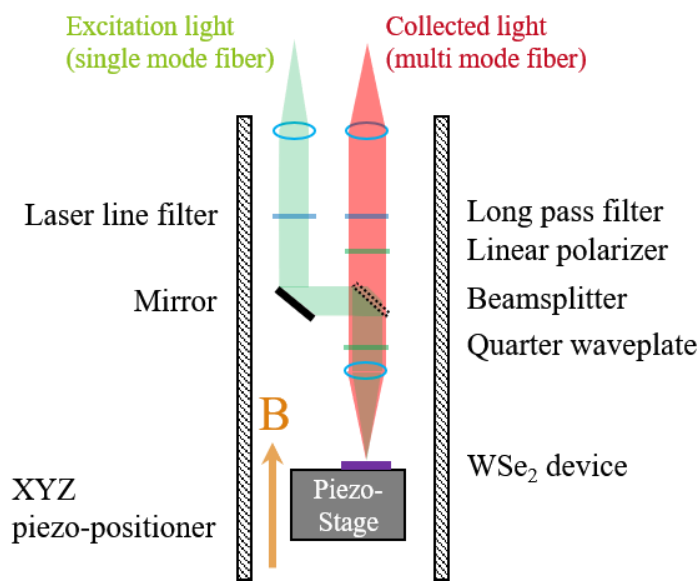

Figure S1. Fiber-based probe setup for photoluminescence (PL) measurement under high magnetic field.

The optical measurements with magnetic field were performed in the National High Magnetic Field Lab (NHMFL) with a 31-Tesla DC magnet and a fiber-based probe (see the schematic in Figure S1). A 532-nm continuous laser was used as the excitation light source. The laser was directed through a single-mode optical fiber and focused by a lens (NA = 0.67) onto the sample. The sample was mounted on a three-dimensional attocube piezoelectric translational stage. The photoluminescence (PL) was collected through a 50/50 beam splitter into a multimode optical fiber, and subsequently measured by a spectrometer (Princeton Instruments, IsoPlane 320) with a CCD camera. A quarter wave plate was used to select the left-handed circularly polarized component of the PL signal.

## 2. Theoretical calculation of exciton Rydberg states in monolayer WSe$_2$

The band structure of monolayer WSe$_2$ near the $K$ (or $K'$) point is nearly parabolic for both the conduction and valence bands[1-3]. The electron and hole effective masses are estimated to be around $0.38m_e$ and $0.44m_e$ ($m_e$ is the free-electron mass), respectively according to the average values of various calculations based on the density-functional theory (DFT) [1-4]. So, the reduced mass ($\mu$) for the exiton in monolayer WSe$_2$ is around $0.2m_e$. The effective electron-hole interaction in two-dimensional (2D) transition-metal dichalcogenides (TMDs) can be approximated by a quasi-2D potential of the following form in the momentum space:[5-10]

$$V(q) = -\frac{e^2}{2A\kappa\varepsilon_0 q(1+q\rho_0)} \tag{S1}$$

Here $A$ is the sample area; $\kappa$ is the effective static dielectric constant of the 2D material; $\rho_0$ is an empirical parameter related to the finite thickness of the 2D material and the screening length of the $q$-dependent dielectric screening[11]. The band-structure effect can lead to another $q$-dependent factor[12], which can also be absorbed into the parameter $\rho_0$. The Fouier transform of $V(q)$ takes the following form in the real space:

$$V(r) = -\frac{e^2}{8\varepsilon_0 r_0}\left[H_0\left(\frac{\kappa r}{r_0}\right) - N_0\left(\frac{\kappa r}{r_0}\right)\right] \tag{S2}$$

Here $r$ is radial coordinate in two dimensions; $H_0$ and $N_0$ denote the Struve and Neuman functions of zeroth order, respectively. In the spherical effective-mass approximation, the Hamiltonian for the electron-hole relative motion in the exciton is:

$$H_X = -\frac{\hbar^2 \nabla^2}{2\mu} + V(r) \tag{S3}$$

The eigenvalues of the low-lying states can be obtained via the Rayleigh-Ritz variational method[13, 14] with a finite basis set of localized functions. The basis functions for the s-like states take the exponential form:

$$f_n(r) = C_n e^{-\alpha_n r} \tag{S4}$$

The exponents $\alpha_n$ follow an even-tempered series $\alpha_n = \alpha g^n$ ($n = 0, ..., N-1$). $C_n = 2\alpha_n/\sqrt{2\pi}$ is the normalization constant. $\alpha$ and $g$ are variational parameters to minimize the eigenvalues of the low-lying states. In our calculation, we choose $\alpha = 0.025$, $g = 1.4$, and $N = 15$. The basis set is then orthogonalized via the Gram-Schmid procedure. We adopt an effective dielectric constant $\kappa = 3.97$. The corresponding atomic unit for exciton energy is $R_X = (\mu/\kappa^2)$ 13.6 eV = 0.172 eV and the atomic unit for distance is $a_X = (\kappa/\mu)$ 0.529 Å = 10.5 Å. We found that, by using $\kappa = 3.97$ and $r_0/\kappa = 1.2\ a_X$, the model potential can well describe the measured energy levels of the five lowest-lying excitonic states in monolayer WSe$_2$. Both parameters are physically reasonable and close to the values considered in Ref. 10.

After digonalizing the Hamiltonian $H_X$ (S3) within the orthogonalized basis set, we obtain the five lowest-lying eigenstates. To check the suitability of the basis set, we also diagonalize the Hamiltonian for the ideal 2D exciton (by setting $\rho_0 = 0$) within the same basis set. The resultant eigenvalues agree with the exact values given by $R_X/(n – 1/2)^2$ with error smaller than $2 \times 10^{-6} R_X$ for the four lowest-lying eigenstates and smaller than $2 \times 10^{-5} R_X$ for the fifth eigenstate.

To describe the diamagnetic shift of the excitonic states, we consider the effect of a constant magnetic field ($B$) perpendicular to the 2D material by the following Hamiltonian:[15]

$$H = \frac{1}{2\mu}(\boldsymbol{p} + e\boldsymbol{A})^2 + V = \frac{1}{2\mu}(p^2 + 2e\boldsymbol{p}\cdot\boldsymbol{A} + e^2 A^2) + V \tag{S5}$$

where $\boldsymbol{A} = \frac{1}{2}\boldsymbol{B}\times\boldsymbol{r}$ (S6)

$$A^2 = \frac{1}{4}r^2 B^2 \tag{S7}$$

$$\boldsymbol{p}\cdot\boldsymbol{A} = \frac{1}{2}\boldsymbol{B}\cdot(\boldsymbol{r}\times\boldsymbol{p}) = \frac{1}{2}\boldsymbol{B}\cdot\boldsymbol{L} \tag{S8}$$

For $s$-like states, the zero angular momentum ($L$) makes $\boldsymbol{p}\cdot\boldsymbol{A} = 0$. The Hamiltonian becomes $H = H_X + H_{DM}$, where $H_{DM}$ is the diamagnetic term:

$$H_{DM} = \frac{e^2}{8\mu} r^2 B^2 \tag{S9}$$

The first-order perturbation theory gives the diamagnetic shift of the exciton energy level:

$$\Delta E_{DM} = \frac{e^2}{8\mu}\langle r^2\rangle B^2 = \sigma B^2 \tag{S10}$$

Here $\langle r^2 \rangle = \langle \psi | r^2 | \psi \rangle$ is the expectation value of $r^2$ over the exciton envelope wave function $\psi$. The exciton's root-mean-square radius is $r = \sqrt{\langle r^2 \rangle} = \sqrt{8\mu\sigma}/e$.

The first-order perturbation theory gives good approximation only in the low field. In the full calculation, we have calculated the exact diamagnetic shift by diagonalizing the total Hamiltonian $H = H_X + H_{DM}$ for each magnetic field. We note that, as the $H_{DM}$ term behaves like a parabolic confining potential, the exciton size will shrink when the magnetic field increases. The calculated exciton sizes in the main paper are the values in the weak field limit.

### 3. Valley polarization of exciton Rydberg states in monolayer WSe$_2$

Compared to prior reflection experiments, our PL experiment has a unique advantage – it can readily reveal the valley polarization of excitons[16-20]. The two electronic valleys in monolayer WSe$_2$ are known to coupled exclusively to light with opposite helicity. For instance, we can generate excitons in one valley with left-handed (L) circular polarization, and detect the exciton population at the same (opposite) valley by PL with left-handed (L) (right-handed, R) circular polarization. The difference between the PL intensity ($I$) of the LL and LR spectra indicates the valley polarization ($\eta$) of the excitons[17, 20], which is defined as $\eta = (I_{LL} - I_{LR})/(I_{LL} + I_{LR})$. Prior studies have compared the valley polarization of 1s and 2s exciton states[21], but further studies for higher excited exciton states are still lacking.

We have measured the helicity-resolved PL spectra of 1s – 4s states in monolayer WSe$_2$ under near-resonance laser excitation (Figure S2). While the 1s state shows only 10% valley polarization, the 2s state shows much higher (64%) valley polarization, and the 3s state shows even higher (71%) valley polarization (the 4s PL is too weak to accurately extract the valley polarization). The valley polarization appears to increase monotonically from low-lying to high-lying states.

The superior valley polarization of the high-lying states to low-lying states can be accounted qualitatively by two competing factors – the exciton intravalley relaxation and intervalley scattering[21]. On the one hand, the exciton intravalley relaxation increases with the number of relaxation channels. The higher states have more relaxation channels than the lower states; they thus have shorter lifetime. On the other hand, the intervalley scattering depends on the electron-hole exchange interaction, which scales linearly with the probability that the electron and hole overlap with each other within the exciton wavefunction[22]. The large-size higher excited states have smaller electron-hole overlapping probability, and hence weaker electron-hole exchange interaction, than the small-size lower states. Overall, the higher states have faster intravalley relaxation and slower intervalley scattering than the lower states. The high-lying excitons have a higher tendency to relax before they can be scattered to the other valley. Therefore, the higher states are expected to have more robust valley polarization than the lower states.

However, we must note that our experiment uses only one excitation laser at 633 nm. The high-lying exciton states are somewhat closer to the excitation photon energy than the low-lying states. The closer resonance favors the valley polarization of the high-lying states. More detailed experiments with tunable laser wavelength are needed to elucidate the valleytronic properties of different exciton Rydberg states.

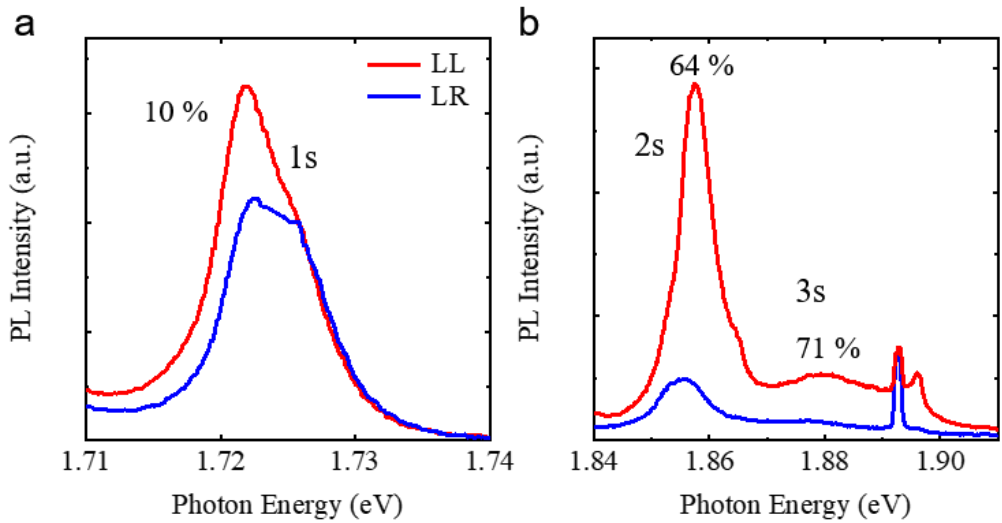

Figure S2. Valley polarization of exciton Rydberg states. (a) The PL spectra of the exciton 1s state in left-left (LL) and left-right (LR) circular polarization measurement geometries. (b) The same PL spectra for the 2s and 3s states. We denote the valley polarization of the 1s, 2s, and 3s states. The excitation source is a 633-nm HeNe laser. The sharp features at 1.893 eV are Raman peaks.

**Supporting References:**